\pdfoutput=1
\documentclass[11pt,a4paper]{article}
\usepackage[T1]{fontenc}
\usepackage[utf8]{inputenc}
\usepackage{lmodern,microtype}
\usepackage[margin=26mm]{geometry}
\usepackage{graphicx,booktabs,tabularx,longtable,enumitem,fancyvrb}
\usepackage{amsmath,amssymb,amsthm}
\usepackage[round,authoryear]{natbib}
\usepackage[hidelinks]{hyperref}
\newtheorem{proposition}{Proposition}
\newcommand{\E}{\mathbb{E}}
\newcommand{\Description}[1]{}
\title{Changing a View Is Not Amending a Record: Timing and Visibility in a Model of Public Correction}
\author{Ziming Liu\\[0.25em]\small The University of Melbourne, Australia}
\date{16 November 2026}
\hypersetup{
  pdftitle={Changing a View Is Not Amending a Record: Timing and Visibility in a Model of Public Correction},
  pdfauthor={Ziming Liu},
  pdfsubject={Computational social science preprint},
  pdfkeywords={public correction, indexed statement amendment, normative feedback, agent-based modelling}
}
\begin{document}
\maketitle
\begin{abstract}
An accusation may remain on the public record after its author has changed their mind. Observers can respond either to what a person says now or to an earlier statement that has never been explicitly amended. We examine the consequences of these two signals with a generative agent-based model that separates private evidence updates, current expression, and revisions linked to a specific declaration. In paired worlds, evidence, prehistory, and opportunities to revisit that declaration remain fixed. When expression and amendment are synchronized and fully observed, the feedback channels produce identical trajectories. With sparse review, record-based feedback leaves 19.44 percentage points more eligible accusations unamended after 24 rounds (paired-world 95\% bootstrap interval 13.89--25.31). The comparison averages 50 eligible worlds, containing 556 agents, from 128 newly generated worlds. Event traces show how a persistent public cue changes subsequent amendment propensity; an exact small-state calculation independently establishes the timing boundary. Reducing record salience can erase or reverse the increment, and a generic held-state process reproduces the dynamics. The findings show when current expression ceases to track correction of a particular public statement.
\end{abstract}
\medskip

\section{Introduction}
Public accusations can outlive the judgements that produced them. A person sees fuller context, revises an assessment, and perhaps expresses a different view---yet the original accusation remains. Someone reading the latest comment may conclude that the author has changed position; someone encountering the old statement may see continued endorsement. These are different public histories, even when the author's private judgement is the same. We ask when the distinction changes the course of public correction.

The distinction is visible in how correction is measured. Fact-checking comments sometimes precede deletion of Facebook reshares \citep{friggeri}; correction bots prompt replies, resharing, blocking, or deletion \citep{burel}. Community Notes can change subsequent diffusion and deletion, with timing shaping the reach of the intervention \citep{chuai}. Each captures an important response, but none alone specifies whether an author has publicly revised a particular earlier claim. Research on self-corrected debunking examines the original sharer as a potential corrector \citep{chen}. Our question concerns the public information left behind when that correction takes place---or does not.

We call an explicit revision linked to an earlier declaration an \emph{indexed statement amendment}; the index identifies the statement in the event history. We distinguish the author's private assessment, their current expression, and the latest indexed stance actually seen by each observer. A \emph{normative cue} is the local pattern of visible stances that influences how an agent chooses to speak. We hold subsequent factual evidence and private updating fixed, then ask whether feedback from current expressions predicts the same first-amendment history as feedback from the old public record.

Private and expressed opinions can diverge in established models \citep{ye,gastner}, and asynchronous updating can change their collective behaviour \citep{alpha,pofj}. Retraction research examines how corrective information affects belief persistence \citep{lacroix}. We build on these ideas by shifting the observable from a current opinion to the amendment of a named earlier statement. A paired intervention changes which public object supplies the norm while retaining the same evidence, private assessment, and opportunities to review the old statement.

This yields three results. A statement-linked first-amendment outcome separates private change, new expression, and repair of the existing record. A synchronization argument identifies when expression-based and record-based feedback agree exactly; controlled simulations and an exact finite-state calculation show how sparse review separates them. Finally, a generic held-state reduction and record-salience controls locate the boundary of that distinction. The contribution is a testable account of public correction as an event in a continuing record, with clear conditions for agreement and divergence between its observable signals.

\section{Public correction and related models}
\subsection{What empirical correction research observes}
Research on factual correction often measures the response of an audience rather than the fate of the original statement. Friggeri et al.\ \citep{friggeri} track Facebook rumor reshares and fact-checking comments, including subsequent deletion. Burel et al.\ \citep{burel} examine visible reactions to automated corrections addressed to 3,898 social-media users. These studies document public responses at scale. A linked amendment poses a narrower question: did the author revise the specific earlier declaration?

Chen et al.\ \citep{chen} study self-corrected debunking in a survey of 1,317 participants during the early COVID-19 outbreak, connecting sharers' psychological responses with corrective action or intention. Their attention to the original author motivates our author-centred outcome. Chuai et al.\ \citep{chuai} analyse Community Notes using platform diffusion histories and a difference-in-differences design. Their results on timing and deletion concern consequences of a correction attached to a post; our outcome is the author's first explicit amendment of that post. The comparison makes these distinct forms of correction measurable within one event history.

Correction also has social costs when it targets accurate information. Stoeckel et al.\ \citep{stoeckel} find that social corrections can reduce perceived accuracy for true as well as false news. We therefore generate fuller evidence that may support or weaken the initial accusation, rather than treating withdrawal as success by definition. Brady et al.\ \citep{brady} show that visible feedback shapes expressions of moral outrage; this supports modelling a social incentive attached to public stance, distinct from factual evidence.

\subsection{Closest dynamical models and the added comparison}
Ye et al.\ \citep{ye} model private and expressed opinions as distinct states, allowing social influence and conformity to operate differently. The concealed voter model \citep{gastner} uses copying, externalization, and internalization to study consensus. The $\alpha$-EPO model \citep{alpha} makes relative updating rates part of the dynamics. These provide a basis for separating private assessment from public expression. Our endpoint, however, is the first amendment of a designated earlier declaration after an identical private evidence update across treatments.

The poster-observer model \citep{pofj} distinguishes posting from observing, while retraction models \citep{lacroix} analyse how correction moves through a network and affects beliefs. Those mechanisms concern who encounters information and how they respond to it. Here factual exposure is controlled. The experiment changes whether neighbours react to fresh expressions or to the last amendment of a previously indexed declaration.

A statement identifier links each amendment to the declaration it changes; the behavioural kernel, by contrast, needs only the most recently held public stance. We compare these representations explicitly, using paired counterfactual worlds to isolate a generative mechanism in the sense of \citet{edmonds}. The distinction separates event attribution from the state that carries social influence.
\section{Model and experimental design}
\subsection{Evidence, private evaluation, and spontaneous prehistory}
Each synthetic world contains an event with hidden state $Z=(X,I)\in\{0,1\}^2$, representing the occurrence of the alleged act and an associated hostile intention. Ground-truth states are drawn uniformly. A single source generates a fragment $C$, a factual-context cue $F$, and an intention-context cue $J$. Its joint likelihood is
\begin{equation}
 K(c,f,j\mid x,i)=.9\,\ell_{.65}(c,x)\ell_{.90}(f,x)\ell_{.80}(j,i)+.1/8,
 \label{eq:likelihood}
\end{equation}
where $\ell_p(u,v)=p$ when $u=v$ and $1-p$ otherwise. The mixture represents a shared uninformative-source branch; the fragment and full context therefore must not be counted as independent sources.

An agent first sees $E_0=C$ and later $E_1=(C,F,J)$. The Bayesian posterior uses the likelihood marginalized over unseen components, then replaces the partial observation when the full context arrives. Agents retain separate evidence accounts. The primary subjective prior has independent marginals $P(X)=P(I)=0.75$, a chosen scenario parameter alongside a uniform world-state generator.

For agent $i$, private evaluation is
\begin{equation}
 m_i=v_{Ni}P_i(X=1,I=0)+v_{Ii}P_i(X=1,I=1),
\end{equation}
with $v_{Ni}\sim U(.10,.35)$ and $v_{Ii}\sim U(.65,.95)$. These ranges define the synthetic agents' valuation heterogeneity. An ordinary expression is $a\in\{-1,0,+1\}$: questioning, no expressed stance, or condemnation. Intrinsic action utility is $a(2m_i-1)-c_i|a|$, where $c_i\sim U(.05,.25)$ applies symmetrically to questioning and condemnation.

The population has 36 agents on an undirected nearest-neighbour ring in the primary assay. Four pre-context rounds generate spontaneous expressions with a norm coefficient of one. A one-round external accusation cue adds $0.35a$ to expressive utility in the second prehistory round; it is not factual evidence. Before $E_1$, the last nonzero expression by each agent is indexed as its focal declaration. No focal statement is selected using the later private update. Agents without a focal statement remain in the population but contribute neither public channel's numerator.

After prehistory, every primary agent receives and accepts $E_1$. Its private evaluation then stays fixed. Normative feedback enters expression choice, while factual updating is held constant across branches so that the subsequent contrast isolates the effect of the public cue.

\subsection{Two public signals and one coherent decision}
Let $G_{ij}$ be adjacency, $d_i=\max(1,\sum_jG_{ij})$, and $z_j$ indicate that author $j$ has a focal declaration. Let $e_j(t)$ denote current expression and $b_{ij}(t)$ the latest indexed stance received by observer $i$ from author $j$. The two signals are
\begin{equation}
\begin{aligned}
 f_i(t)&=\frac{1}{d_i}\sum_jG_{ij}z_j e_j(t-1),\\
 r_i(t)&=\frac{1}{d_i}\sum_jG_{ij}z_j b_{ij}(t).
\end{aligned}
 \label{eq:signals}
\end{equation}
The actual implementation maintains recipient-specific delivered flow when corrections have incomplete reach. Equation~\eqref{eq:signals} gives the primary full-delivery case. The denominator is full graph degree in both channels, not the number of indexed authors. This preserves identical dilution by non-indexed neighbours.

With record mixture $w\in[0,1]$, the norm is $n_i(t)=(1-w)f_i(t)+wr_i(t)$. Public action utility and choice probability are
\begin{align}
 U_i(a,t)&=a\{2m_i-1+\gamma n_i(t)\}-c_i|a|,\\
 P_i(a,t)&=\frac{\exp[\beta U_i(a,t)]}{\sum_{a'=-1}^{1}\exp[\beta U_i(a',t)]}.
 \label{eq:choice}
\end{align}
The primary parameters are $\gamma=2$, $\beta=3$. Controls include no normative influence ($\gamma=0$), weaker influence ($\gamma=1$), and a mixed channel $w=0.5$.

Each round provides a new expression and an independent Bernoulli review opportunity $Q_i(t)$ with probability $q$ for the focal record. There is \emph{one} stance draw: if review occurs, that same draw determines the new indexed stance. No additional switching penalty is imposed. If the indexed stance changes, an explicit amendment names the current target, and the new statement becomes the next revision target. Each target can be amended only once, but an amendment can itself be amended later. Ordinary silence or a new contrary expression does not retrospectively repair the indexed statement.

A change from condemnation to a neutral indexed stance counts as an explicit withdrawal, not as exoneration. This convention is distinct from ordinary silence, which creates no amendment event. Neighbours receive an emitted amendment at the beginning of the next round. At primary coverage one, all graph neighbours receive it; an amendment emitted in the last observed round is not automatically delivered beyond the observation window. Initial signals and the set of contributing authors are matched between channels.

\subsection{Estimands, risk set, and paired worlds}
The strict risk set in world $s$ contains agents with a focal condemnation, an intrinsically optimal condemnation before $E_1$, and a different intrinsically optimal action after accepting $E_1$. ``Private reversal'' therefore means a change in this decision ranking, not proof of the target's innocence or of a universal moral threshold. Every social branch uses the same risk set, event, graph, values, prehistory, and random streams.

For an eligible agent, $T_i$ is the first amendment round, indexed from zero; $T_i=\infty$ denotes no amendment within the run. At horizon $H=24$, the main outcome is the within-world proportion with $T_i\ge H$. The delay measure is $\min(T_i,H)$, so amendment in the first round has delay zero. We average world-specific outcomes equally and compare record minus flow within each world. A separate all-initial-condemner analysis includes every focal condemnation, regardless of subsequent private change. Worlds with no eligible agents are recorded as undefined for the strict estimand, not as zero delay.

We construct 10,000 paired-world bootstrap samples, preserving each world's record--flow pairing. The resulting intervals reflect variation under the declared synthetic generator; the study's assumptions and parameter sensitivity are assessed separately.

\subsection{Frozen closure experiment and controls}
The primary experiment uses 128 new world seeds, 7100--7227. Each world is paired across six arms: no norm; flow and record at $\gamma=1$; flow, mixed, and record at $\gamma=2$. A synchronized $q=1$ comparison uses those same worlds. Review rates .25 and .75 use their first 64 seeds; sixteen targeted sensitivities use a separate block, 7300--7347, with three arms each. The design yields 3,840 world--arm trajectories, clustered by their shared world seeds.

The sensitivity set varies choice responsiveness, subjective priors, population size, network structure, review and delivery, prehistory norms, record salience, and observation horizon. Settings and the exact small-state audit were recorded before their respective runs. Exploratory results remain archived; the new primary estimate uses the frozen confirmation design.

\subsection{Research assistance and scope of the data}
ChatGPT assisted with literature searches and comparisons, model and experimental-design development, code generation and debugging, verification and analysis scripts, mathematical exposition, and manuscript drafting and revision. Simulation choices were generated by the stated probabilistic model; figures were plotted from archived numerical outputs. Independent code paths and new random seeds were used to check the results. The study involves synthetic agents and no human participants or identifiable platform-user data. The human author is responsible for the design, analysis, and final manuscript.
\section{Analytical boundaries and the strongest rival}
\begin{proposition}[Synchronized equivalence]
With matched initial cues and source sets, coherent stance choices, zero extra switching cost, complete delivery, no record attenuation, and $q=1$, flow-based and record-based branches are pathwise identical under shared random draws.
\end{proposition}
\begin{proof}
Initially the two signals coincide by construction. If they coincide before a round, utilities and coupled stance draws coincide. Every indexed author reviews; its indexed stance consequently equals its current expression, including a neutral stance. Complete delivery makes both next-round signals identical. Induction yields equality of all subsequent actions and first-amendment events. Authors without indexed records contribute to neither numerator.
\end{proof}

\begin{proposition}[Label-free reduction]
Under the primary decision rule, the choice process and first-amendment times can be reproduced by a generic held-state process with no statement identifiers. With full delivery, its owner-level state satisfies
\begin{equation}
 b_j(t+1)=(1-Q_j(t))b_j(t)+Q_j(t)a_j(t).
 \label{eq:hold}
\end{equation}
With incomplete delivery, an observer-specific cache receives changed held states on the same delivery schedule.
\end{proposition}
\begin{proof}
The normative utility depends on cached stances, not on identifier strings. Equation~\eqref{eq:hold} gives exactly the owner-state update of the indexed process. Copying changed states along the same edges reconstructs the observer caches. A first departure from the original held stance is exactly a first amendment. Identifiers recover which declaration was amended, but are unnecessary for this transition kernel.
\end{proof}

The reduction separates two roles of a statement index. An identifier tells us which public declaration was amended; the held stance supplies the normative signal governing the next choice. Two histories with identical private assessment and current expression can therefore predict different amendments if observers retain different old public stances. The held public state is sufficient to express that difference.

Without normative feedback, intrinsic choice probabilities remain fixed. For an initially condemning focal author, after $k$ completed rounds the no-amendment probability is
\begin{equation}
 S_i(k)=\left[1-q\{1-P_i(+1)\}\right]^k.
 \label{eq:survival}
\end{equation}
A round provides a Bernoulli review opportunity, so $k$ counts completed rounds, regardless of how many opportunities were realized. Equation~\eqref{eq:survival} gives the baseline persistence caused by opportunities alone. For bounded utilities, finite $\beta$, and positive $q$, the probability of leaving condemnation stays positive: persistence over a finite observation window is different from an absorbing refusal to amend.

\subsection{From a public cue to amendment propensity}
For an initially condemning agent whose original statement is still unamended, the next-round first-amendment probability conditional on the current history is
\begin{equation}
 h_i(t)=q\{1-P_i(+1,t)\}.
 \label{eq:hazard}
\end{equation}
Review and choice are independent conditional on that history. Holding private evaluation and cost fixed, the logit rule gives
\begin{equation}
 \frac{\partial P_i(+1,t)}{\partial n_i(t)}
 =\beta\gamma P_i(+1,t)\{1-\E_i[a\mid t]\}\geq0.
 \label{eq:derivative}
\end{equation}
For $\beta,\gamma\geq0$, a more condemning observed norm weakly reduces the conditional probability of amendment. Whether old-record or current-expression feedback generates the stronger norm is determined by their evolving histories. An author can express a new, non-condemning stance on a round without review, leaving a positive indexed record in place. Neighbours observing that held stance may in turn sustain condemnation and slow the amendment of their own records. Both channels use the same review opportunities, making this feedback difference measurable.
\section{Results}
\subsection{Fresh worlds confirm an additional finite correction delay}
Of the 128 primary worlds, 50 contain strict-risk agents: 556 eligible agents in total. The broader cohort contains 2,733 initial condemners across all 128 worlds. Table~\ref{tab:main} reports world-equally-weighted outcomes; pooling agents would answer a different question.

\begin{table}[htbp]
\centering\small
\caption{Primary $q=.5$ results after 24 rounds. Complete evidence and correction delivery; zero extra switching cost.}
\label{tab:main}
\begin{tabular}{lrrr}\toprule
Channel & Unamended (\%) & Delay & Current $+1$ (\%)\\\midrule
No norm & 0.00 & 1.415 & 15.02\\
Expression & 29.83 & 12.447 & 47.57\\
Record & 49.27 & 15.871 & 59.20\\\bottomrule
\end{tabular}
\end{table}

\begin{figure}[!htbp]\centering
\includegraphics[width=.74\linewidth]{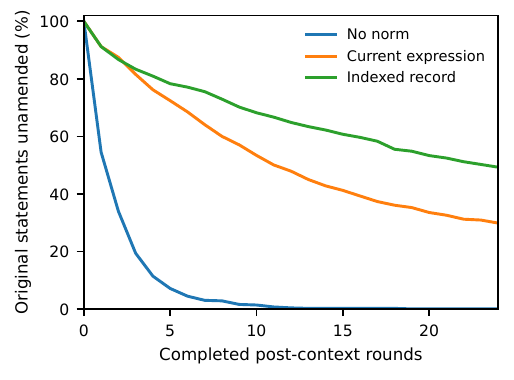}
\caption{Current-expression and indexed-record feedback separate under sparse review. Lines give world-equal means over the same 50 eligible worlds after complete-context evidence is accepted; the initial value precedes post-context decisions.}
\Description{Three curves show first-amendment survival over 24 rounds. No normative influence approaches zero quickly; expression and record feedback retain more unamended statements, with records higher at the endpoint.}
\label{fig:survival}
\end{figure}

Record feedback increases the strict never-amended fraction by 19.44 percentage points (95\% interval 13.89--25.31) and truncated delay by 3.424 rounds (2.560--4.288) relative to flow. In the broader cohort, the corresponding never-amended increment is 9.00 points (6.28--11.94). Figure~\ref{fig:survival} shows the full first-amendment trajectory.

A zero never-amended fraction under no normative feedback coexists with 15.02\% current condemnation at the last round. An initial statement can be amended and its author can later condemn again. Tracking first amendment and current expression separately prevents the former from being mistaken for lasting consensus.

\subsection{Archived trajectories explain how the gap accumulates}
We trace the process behind the endpoint contrast using the frozen primary trajectories. The same strict cohort is followed across both channels, with identical private states and review schedules. The following process measures were calculated after the primary analysis and are descriptive diagnostics.

\begin{figure}[tb]
\centering\includegraphics[width=.74\linewidth]{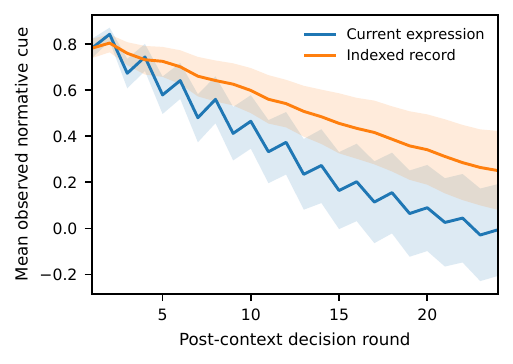}
\caption{The initially matched normative cues separate as new expressions and indexed statements are refreshed at different rates. Fixed strict cohort; world-equal means with pointwise bootstrap intervals. The early crossing remains visible.}
\Description{Two initially coincident lines show mean normative cues over 24 decision rounds. The current-expression line is briefly higher, then falls faster than the indexed-record line. Both uncertainty bands widen over time.}
\label{fig:mechanism}
\end{figure}

The two channels begin with identical private states, public cues, review opportunities, and first-round actions. Their first-round mean norm is 0.781 in both, and mean prospective amendment probability $q(1-P_i(+1,t))$ is 0.0798. There is therefore no initial norm advantage inserted into the record branch. Figure~\ref{fig:mechanism} shows that subsequent cues separate, with an early crossing rather than uniform record dominance. By decision round 8, the mean cue is 0.559 under flow and 0.641 under records; corresponding prospective probabilities are 0.1286 and 0.1090. At round 24, the cues are $-0.006$ and 0.251, and probabilities 0.2619 and 0.2058. These probabilities describe the same fixed cohort; they are not empirical hazards estimated only among still-unamended agents.

Let $A_{s,c,t}$ be the proportion of the fixed strict cohort in world $s$ whose \emph{first} amendment occurs at round $t$ in channel $c$. Because these events are mutually exclusive for each original statement,
\begin{equation}
 \overline S_R(H)-\overline S_F(H)
 =\sum_{t=0}^{H-1}\overline{A_{F,t}-A_{R,t}}.
 \label{eq:accounting}
\end{equation}
The event log attributes 4.74 percentage points of the final gap to differences in first amendments during rounds 1--4, 12.20 points during rounds 5--12, and 2.49 points during rounds 13--24 (unrounded sum 19.4403). This is an accounting decomposition by event time. The controlled contrast, together with Equation~\eqref{eq:derivative} and the trajectories, connects persistent cues to reduced amendment propensity within the specified model.

\subsection{Synchronization eliminates the incremental record effect}
When every author can review in every round ($q=1$), current-expression and record-based feedback produce identical state and amendment histories in all 128 paired worlds. Both yield 26.83\% never-amended after 24 rounds. This exact agreement is the empirical counterpart of the synchronization proposition: the additional record delay vanishes even though normative persistence remains.

With review probability $q=.25$, the paired record increment is 30.54 points (21.04--40.00), compared with 12.11 points (5.10--20.33) at $q=.75$; each contrast uses 27 eligible worlds from its specified 64-world block. The mixed flow--record condition in the primary sample has 49.45\% unamended, marginally above the pure-record condition. Together these cases delimit the observed clock effect without implying a globally monotone response to record weight.

At $\gamma=1$, the record--flow increment is 4.32 points (2.08--7.06), versus 19.44 points at $\gamma=2$. The paired difference of these increments is 15.12 points (8.94--21.41). The larger delay is not only the mechanical effect of fewer personal review opportunities: review opportunities are shared between channels, and increasing normative strength changes their difference.

\subsection{An exact small-state model and a generic cache reproduce the mechanism}
A separate two-agent construction fixes post-context evaluation at $m=.25$, cost at .1, and $\beta=3$. Both original records are positive, despite the subsequent intrinsic preference for questioning. Before the focal author's first amendment, its record remains positive. A transient state needs only the two current expressions and the other author's held record: $3^3=27$ states. Enumerating the substochastic transition matrix gives the exact survival probability, without simulation sampling error.

\begin{figure}[!htbp]\centering
\includegraphics[width=.74\linewidth]{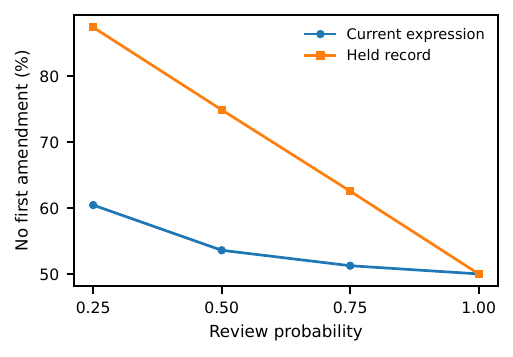}
\caption{Exact enumeration confirms separation and its synchronization boundary in a two-agent fixture. The difference is computed from a 27-state transient matrix, not fitted to the population simulation. Both original records are positive, with fixed post-context private evaluation $m=.25$.}
\Description{An exact two-agent calculation shows the two channels separating when review is less frequent and meeting when review occurs every round.}
\label{fig:exact}
\end{figure}

At $q=.5$, $\gamma=2$, no-amendment probability at round 24 is 53.62\% under flow and 74.91\% under held-record influence. At $q=1$ the curves coincide at every enumerated time. At $\gamma=0$ they coincide with Equation~\eqref{eq:survival}. Figure~\ref{fig:exact} illustrates this independently specified fixture; its percentages are not substitutes for the larger-population estimates.

A separately written reference program, without the production simulator or statement identifiers, reproduced all 3,840 archived trajectories. The audit covered 3,815,424 agent-rounds and 487,225 amendments, with a maximum numerical difference of $6.66\times10^{-16}$. This establishes that held-state feedback suffices for the simulated choices; identifiers allow the experiment to track amendment targets and audiences.

\subsection{Salience reversals constrain the interpretation}
The baseline retains the last observed indexed stance without attenuation. A targeted control assigns it weight $2^{-\mathrm{age}/h}$, where age is measured from the statement's emission and $h$ is a half-life. The graph-degree denominator stays unchanged, so attenuation can reduce total normative input as well as reweight older and newer information. A second control normalizes the weights to preserve total indexed-source weight and thus isolates relative age weighting more closely.

With record half-life four, the unnormalized record--flow contrast reverses to $-32.79$ points ($-49.22$ to $-17.71$); normalized weights give $-1.56$ ($-14.01$ to 10.08). At half-life twelve, the corresponding estimates are $-26.23$ ($-39.11$ to $-13.78$) and 8.45 ($-3.30$ to 19.70). The normalized intervals cross zero. The comparison shows that attenuation changes both the age composition of the record and, without normalization, the total weight of social feedback.

\begin{figure}[!htbp]\centering
\includegraphics[width=.74\linewidth]{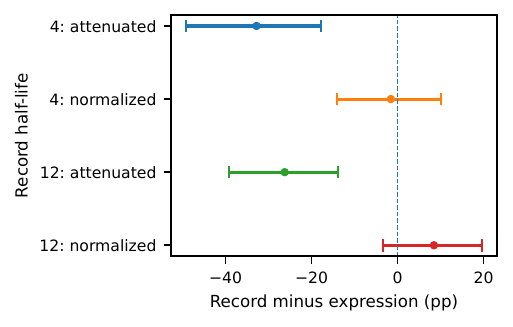}
\caption{Apparent memory reversals depend on how total normative weight is treated. Each estimate is record minus flow within the same 19 eligible worlds from 48 generated worlds. Error bars are pointwise 95\% paired-world bootstrap intervals. Normalization preserves indexed-source weight, not any empirically measured human attention budget.}
\Description{Four interval estimates contrast decaying-record signals with current expression. Unnormalized weights yield negative differences; normalized contrasts are closer to zero.}
\label{fig:memory}
\end{figure}

\subsection{Other boundaries and non-identification cases}
\begin{figure}[!htbp]\centering
\includegraphics[width=.74\linewidth]{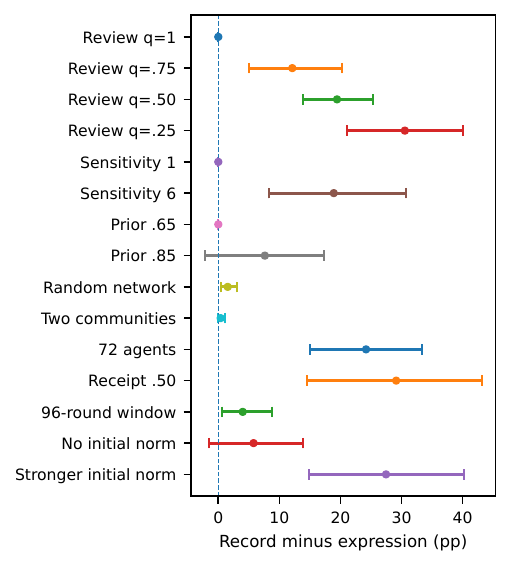}
\caption{The incremental record effect is bounded rather than universal. Pointwise 95\% intervals accompany within-scenario paired differences. Sample sizes differ: 50 eligible worlds for primary/synchronized, 27 for the .25/.75 clocks, generally 19 for sensitivities, three for priors .65, and 35 for priors .85. The neutral-prior case has no strict-risk worlds and is omitted as undefined. The 96-round endpoint uses its stated longer horizon.}
\Description{Fifteen scenario-specific interval estimates show positive, zero, and imprecise record-minus-expression differences; the synchronous scenario is exactly zero.}
\label{fig:boundaries}
\end{figure}

Figure~\ref{fig:boundaries} collects targeted sensitivities. Doubling population size retains a positive increment, while denser random networks and two-community networks yield much smaller differences. Because these changes also affect density, they are interpreted as scenario contrasts rather than a topology-only effect. Low choice sensitivity yields no strict never-amended difference at round 24. The higher-prior interval includes zero; the lower-prior condition has only three eligible worlds. With a neutral prior, no worlds enter the strict risk set, leaving that conditional outcome undefined.

With a 96-round observation window, the paired increment narrows to 3.99 points (0.66--8.86); current-expression and record means are 14.22\% and 18.21\%, respectively. That sensitivity uses its own seed block. When only half of amendments are delivered, the increment is larger because some observers retain older cues. Delivery and intrinsic review timing therefore offer distinct routes to public-record persistence.

\section{Discussion}
\subsection{Public correction requires more than a new expression}
The central result concerns the fate of an earlier accusation after fuller evidence changes an author's private preference. Even with the same chance to review that accusation, different public signals can create different feedback. A new expression immediately enters the current-expression channel; the indexed-record channel changes when the original statement is amended and the update reaches observers. The resulting divergence feeds back into later expressive choices and the timing of other authors' amendments.

This mechanism gives a specific interpretation to the varied outcomes in empirical correction research. A fact-check attached to a post, deletion, reduced diffusion, a self-reported change, and an explicit author amendment answer different questions \citep{friggeri,burel,chen,chuai}. By observing an identified old statement, its amendment event, and the audience's receipt, the model makes visible a gap that a latest-post measure can conceal: the author may change what they say today while yesterday's accusation remains legible to others.

\subsection{What is new and what is shared with earlier models}
The result locates a public-correction problem inside a familiar family of social-interaction models. Earlier work separates private and expressed opinions or lets them update at different rates \citep{ye,gastner,alpha,pofj}. Here we compare two public objects after the same private update. Synchronization supplies an exact equality condition, while sparse review allows the objects to diverge and alter first-amendment outcomes. The size and direction of that increment depend on how the record is maintained and how strongly its contents influence later expression.

The generic-cache reduction clarifies which part of the representation matters. The index makes an amendment attributable to a particular declaration. The cached stance produces the dynamics. These are complementary roles: event attribution defines what counts as public correction, while a held-state model explains how the visible record can outlast a changed current view. The model's controlled counterexamples give that relationship an observable boundary.

\subsection{Implications and limits}
An empirical study could follow authors from a private evidence update through their next expression, their opportunities to revisit a particular post, any linked amendment, and what the original audience receives. Interventions that link an updated stance to an earlier declaration would then address the public-record channel directly. The timing sensitivity of correction responses reported by Burel et al.\ \citep{burel} and Chuai et al.\ \citep{chuai} makes these distinct observables worth collecting. As Stoeckel et al.\ \citep{stoeckel} show, inaccurate corrections are also possible; any application would need to track the underlying factual evidence.

Several assumptions govern the scope of the result. The priors and choice coefficients are theoretical inputs rather than human estimates, and the primary subjective prior differs from the world generator. Review opportunities follow an external clock; the model does not explain why individuals return to old posts. Symbolic evidence replaces natural-language interpretation, and each author has one focal declaration rather than a feed of overlapping disputes. Norms affect expression but leave post-context private beliefs fixed. Some network contrasts vary density along with structure. Time is measured in model rounds. Finally, the endpoint is first amendment: authors may later condemn again, and the experiment does not measure reputational repair or lasting agreement.

\section{Reproducibility}
The experiment specification, source snapshot, seeds, raw trajectories, sensitivity analyses, and reference implementation are retained together. The primary worlds use seeds set apart from earlier exploration. Changes in this textual revision affect exposition and the archived-trajectory diagnostics; the original simulated events and primary estimates remain unchanged.

The research archive includes executable analysis commands, model code, stored closure trajectories, a separate reference implementation, and an ODD-style description \citep{odd}. Reanalysis recovered all 3,840 first-amendment endpoints and the paired primary estimate with zero mismatches. The reference implementation independently reconstructed the underlying trajectories. Public release and persistent archival identifiers will be specified when the reproducibility materials are deposited.

\section{Conclusion}
A public accusation can remain visible after its author's current judgement has changed. In a generative social model with identical evidence and review opportunities, feedback from current expression and from an indexed old declaration coincide when updates are synchronized. Sparse review separates these signals and, in the primary setting, increases the share of accusations never explicitly amended by 19.44 percentage points over 24 rounds. The difference shrinks or reverses under alternative record-salience settings, and a generic held-state process accounts for the choice dynamics. These results make explicit amendment---rather than opinion change alone---an informative endpoint for studying how public correction unfolds.

\bibliographystyle{plainnat}
\bibliography{references}
\clearpage
\appendix
\section*{Supplementary Material (V3.1)}
\addcontentsline{toc}{section}{Supplementary Material (V3.1)}
\section{Purpose, patterns, and scope}
This supplement specifies the generative assay in the seven-part ODD reporting structure \citep{odd}. The unit of analysis is an indexed declaration that can be amended after an agent accepts fuller evidence. We test its amendment history under two public cues: neighbours' current expressions and neighbours' most recently observed indexed statements. Verification targets are synchronization equality, divergence under sparse review, the analytic no-norm baseline, and sensitivity to the visibility of older statements.

\section{Entities, state variables, and scales}
There are three types of objects: a synthetic event, individual agents, and public declarations/amendments. The event contains an unobserved binary factual state and binary intention state, plus a jointly generated fragment and full-context bundle. Agents have individual evidence accounts, a four-state posterior, two evaluation values, expressive cost, a current expression, and a locally observed cache of neighbours' indexed stances. Indexed declarations have an author, original emission time, current target identifier, current owner stance, and links from amendments to the target they revise.

The default population is 36 on a degree-two ring. A population-72 variant retains degree two. The random graph independently includes each undirected edge with probability .30. The two-community graph splits agents into equal halves, with edge probabilities .65 within and .10 between halves. There are no self-loops. Degree-zero nodes use denominator one. Both alternative network families are substantially denser than the ring; comparisons therefore combine density and organization.

Time is discrete and dimensionless. Four prehistory rounds precede full context. The principal observation window then contains 24 rounds; one sensitivity extends it to 96. An agent's first post-context amendment is coded zero, so the truncated delay is zero for an immediate amendment and $H$ for no amendment within the window. A model round is not identified with an hour, day, or platform visit.

\section{Process overview and scheduling}
Initialization samples the event, source outcome, individual values, graph, and subjective priors. Agents see the fragment. Four simultaneous rounds generate the public history. Each round computes norms from the previous expressions, draws one expression per agent, and updates the last nonzero focal declaration. An external accusation cue is present only in the second prehistory round and is separate from evidence. At the end of prehistory, focal declarations and their audience caches are fixed before branch comparisons.

The full evidence bundle is delivered at the beginning of the assay. Every primary agent accepts it through the same-source Bayesian account. Private quantities and the strict risk set are computed once. Each subsequent round then executes:
\begin{enumerate}[itemsep=1pt]
\item Deliver the previous round's amendments to their selected receivers; update those caches and timestamps.
\item Compute flow and record signals using matched source sets and graph-degree denominators.
\item Compute the three public-choice probabilities and draw one stance per agent.
\item Draw the Bernoulli review opportunity independently of stance selection; non-indexed agents cannot review an indexed statement.
\item For reviewing authors, use the same stance draw to update the owner record. If it changes, emit an amendment pointing to the current target and create the successor target.
\item Select amendment recipients, compute the recipient-specific flow for the next round, and save individual states, events, and endpoints.
\end{enumerate}
An unchanged owner stance produces no amendment and does not refresh its public timestamp. An ordinary expression produces no link to an earlier statement. Last-round amendments can count as author actions but have not yet reached recipients inside the observation window. All primary update rules are simultaneous at the within-round decision stage; their causal ordering is encoded in these phases.

\section{Design concepts and behavioural assumptions}
First-amendment times emerge from independent review opportunities and local feedback operating through probabilistic expressive choices. Condemnation can be issued, retained, amended, and issued again under the same action rule. Each agent's private evaluation is fixed after accepting full context, which makes subsequent treatment differences attributable to the public-signal channel and its coupled choices.

Each agent observes cues delivered along its network edges and retains the last indexed stance received from each neighbour. The model has no institutional sanction decision or inference about a neighbour's sincerity. Review opportunities follow an externally assigned Bernoulli schedule. Separate keyed streams drive choices, reviews, graph edges, and amendment delivery; paired treatment arms reuse their corresponding streams.

The normative coefficient governs how much locally observed public expression shifts action utility. Its values are theoretical design parameters, varied in the sensitivity experiments. They specify a social-interaction mechanism rather than an estimated human behavioural coefficient.

\section{Initialization and exact support checks}
The hidden event state is uniform on four combinations. Source likelihood is the .9-valid/.1-uninformative mixture in the main paper. The positive-fragment marginal probability under a uniform prior yields factual posterior .635; intent remains .5. The maximum initial private evaluation under the specified value support is then .41275. Because condemnation must exceed silence, its necessary threshold is $(1+c)/2$, at least .525. Thus a neutral-prior branch has no strict sincere-reversal risk set in this setup.

With subjective marginals .75, a positive fragment can make condemnation optimal before full context reverses the ranking. Exact support checks enumerate 16 prior pairs, eight source outcomes, and eight value/cost endpoints (1,024 cells); the resulting table is retained in \texttt{results/r1\_end\_to\_end/EXACT\_REACHABILITY.csv}. The closure also includes neutral, .65, and .85 prior conditions. No neutral-prior world reaches the strict risk set; the .65 condition yields only three eligible worlds.

The focal declaration is the last nonzero pre-context expression. It is not chosen because the subsequent result supports the desired phenomenon. A nonzero expression can be intrinsically disfavoured because public choice is stochastic or norm-influenced. Such statements remain in the broader all-initial-condemner cohort, but do not automatically enter the sincere-reversal subgroup.

\section{Input data and experimental manifest}
Inputs are synthetic event outcomes, the graph generator, agent parameter supports, and fixed random seeds. The closure contract was frozen before the new runs. All worlds remain in the archive, including those whose full evidence supports the accusation or produces no strict-risk reversal.
\begin{table}[htbp]
\centering
\caption{Frozen execution manifest. Scenario--world pairs reuse seeds; they are not independent across scenarios.}
\begin{tabular}{lrrr}
\toprule Scenario & Seeds & Worlds & Arms\\\midrule
Primary & 7100--7227 & 128 & 6\\
Synchronized review & 7100--7227 & 128 & 3\\
Review probability .25 & 7100--7163 & 64 & 3\\
Review probability .75 & 7100--7163 & 64 & 3\\
Choice sensitivity 1 & 7300--7347 & 48 & 3\\
Choice sensitivity 6 & 7300--7347 & 48 & 3\\
Subjective priors .65 & 7300--7347 & 48 & 3\\
Subjective priors .85 & 7300--7347 & 48 & 3\\
Neutral subjective priors & 7300--7347 & 48 & 3\\
Random graph & 7300--7347 & 48 & 3\\
Two communities & 7300--7347 & 48 & 3\\
Population 72 & 7300--7347 & 48 & 3\\
Salience half-life 4 & 7300--7347 & 48 & 3\\
Salience half-life 12 & 7300--7347 & 48 & 3\\
Half-life 4, normalized & 7300--7347 & 48 & 3\\
Half-life 12, normalized & 7300--7347 & 48 & 3\\
Correction receipt .5 & 7300--7347 & 48 & 3\\
Horizon 96 & 7300--7347 & 48 & 3\\
No prehistory norm & 7300--7347 & 48 & 3\\
Prehistory norm 2 & 7300--7347 & 48 & 3\\
\bottomrule
\end{tabular}
\end{table}

The manifest totals 1,152 scenario--world pairs and 3,840 world--arm trajectories. The principal six-arm factorial and synchronized three-arm experiment use identical primary seeds; the two extra review rates use the first 64 of those seeds. Targeted sensitivities share the separate 48-seed block. Reusing seeds improves pairing within comparisons but does not turn these repeats into new independent worlds. There are 176 distinct seeds overall. The exact two-agent assay is a separate mathematical fixture, not another random-world sample.

\section{Submodels and strong baselines}
\subsection{Bayesian evidence}
For observed components $O$, the source likelihood is the sum of the joint likelihood over outcomes consistent with $O$. The posterior is proportional to that marginalized likelihood times the subjective prior. Full context replaces the partial observation of the same source; the old marginal likelihood is not multiplied by the complete one. Factual and intention marginals and private evaluation are exported separately. Sharing a source family does not mean sharing a mutable posterior account.

\subsection{Expression and correction}
Public utility is linear in stance and the received normative cue, with the same expression cost for questioning and condemnation. A finite-temperature logit maps these utilities to the three possible actions. Each agent draws one stance per round; when a review opportunity occurs, that draw also updates its held statement. A review that repeats the held stance produces no amendment. Changing it creates a linked amendment, which can itself become the target of a later amendment. All closure arms use zero additional switching cost.

\subsection{Visibility attenuation and its normalization}
With half-life $h$, received record $j$ has weight
\[
 v_{ij}(t)=G_{ij}z_j2^{-\max(0,t-1-\tau_{ij})/h},
\]
where $\tau_{ij}$ is the emission time of the last received indexed stance. The attenuated signal is $\sum_jv_{ij}b_{ij}/d_i$. Its normalized counterpart divides this signal by $\sum_jv_{ij}/\sum_jG_{ij}z_j$ when the latter is positive. With no indexed neighbours the signal is zero. This preserves the baseline total indexed-neighbour weight relative to full degree while changing relative age weights. It is a diagnostic normalization, not an empirical attention law. A failed or unchanged review does not reset a timestamp.

\subsection{Independent finite-state calculation}
For the two-agent fixture, let $e_1,e_2,b_2\in\{-1,0,+1\}$. Conditional on the focal agent not yet amending, $b_1=+1$. This gives 27 transient states. In each state, the flow norms are $(e_2,e_1)$ and record norms $(b_2,+1)$. The joint stance probability is the product of the two logit probabilities. If the focal draw is not $+1$, survival requires that its review not occur, multiplying the transition by $1-q$. The other agent's held state either remains $b_2$ with probability $1-q$ or becomes its current draw with probability $q$. Summing over draws yields the substochastic matrix $Q$. With $v_0$ concentrated on $(+1,+1,+1)$, survival after $k$ rounds is $v_0Q^k\mathbf 1$.

The separate exact-calculation script enumerates $\gamma\in\{0,1,2\}$ and $q\in\{.25,.5,.75,1\}$ through round 96. It checks pathwise equality at $q=1$, geometric survival at $\gamma=0$, nonnegative transitions, and row sums no greater than one. This two-agent fixture supplies a finite-state mechanism check alongside the sampled population experiment.

\subsection{Generic cache and identification}
The held-state kernel stores each owner's current indexed stance, the copies last received by observers, their timestamps, and amendment delivery outcomes. Given the same choice and review draws, it reproduces the indexed-declaration trajectories. Statement identifiers link events to targets, while transition probabilities depend only on cached stances. The simulation therefore isolates stale public-state feedback rather than a separate psychological preference for acknowledging one's past claims.

A smaller state with only private evaluation and current expression loses information about earlier public records. Consider two otherwise identical histories with all-positive versus all-negative neighbour caches. For $w\gamma>0$, their next-round choice probabilities differ despite identical current expressions. The cached public state is therefore needed to predict this model's future amendments.

\section{Complete sensitivity contrasts and uncertainty}
\begin{table}[htbp]
\centering
\caption{All frozen record-minus-flow contrasts in the strict-risk fraction never amended by the configured horizon. Intervals are pointwise; units are percentage points.}
\begin{tabular}{lrrr}
\toprule Condition & Eligible worlds & Difference & 95\% interval\\\midrule
Primary & 50 & 19.44 & [13.89, 25.31]\\
Synchronized review & 50 & 0.00 & [0.00, 0.00]\\
Review probability .25 & 27 & 30.54 & [21.04, 40.00]\\
Review probability .75 & 27 & 12.11 & [5.10, 20.33]\\
Choice sensitivity 1 & 19 & 0.00 & [0.00, 0.00]\\
Choice sensitivity 6 & 19 & 18.90 & [8.25, 30.73]\\
Subjective priors .65 & 3 & 0.00 & [0.00, 0.00]\\
Subjective priors .85 & 35 & 7.62 & [-2.22, 17.30]\\
Neutral subjective priors & 0 & Undefined & ---\\
Random graph & 19 & 1.54 & [0.38, 3.01]\\
Two communities & 19 & 0.38 & [0.00, 1.13]\\
Population 72 & 19 & 24.19 & [14.94, 33.39]\\
Salience half-life 4 & 19 & -32.79 & [-49.22, -17.71]\\
Salience half-life 12 & 19 & -26.23 & [-39.11, -13.78]\\
Half-life 4, normalized & 19 & -1.56 & [-14.01, 10.08]\\
Half-life 12, normalized & 19 & 8.45 & [-3.30, 19.70]\\
Correction receipt .5 & 19 & 29.12 & [14.49, 43.14]\\
Horizon 96 & 19 & 3.99 & [0.66, 8.86]\\
No prehistory norm & 19 & 5.77 & [-1.52, 13.88]\\
Prehistory norm 2 & 19 & 27.46 & [14.91, 40.28]\\
\bottomrule
\end{tabular}
\end{table}

The principal bootstrap resamples eligible worlds, retaining the record--flow pairing within each draw. The broader analysis includes worlds with any initial condemnations. Worlds with no strict-risk agents have an undefined strict-cohort outcome and are excluded only from that conditional mean; the number excluded is reported. Percentile intervals use 10,000 draws and analysis seed 91001. They are pointwise intervals across simulated worlds; synchronized conditions yield exactly zero because the paired histories coincide.

Only within-scenario contrasts support controlled channel comparisons. Comparing two sensitivity means from different risk-set distributions does not isolate a parameter's unconditional population effect. The difference of record--flow contrasts across $\gamma=1$ and $2$ is evaluated on the same 50 eligible primary worlds. The mixed-channel arm is reported rather than discarded even though it does not show a strictly ordered dose response.

\section{Verification, lineage, and limitations of review}
The source and results archive remains frozen. The \texttt{closure\_model.py} variant adds salience options; tests with attenuation disabled match the historical model. All 174 tests in the full working project and 39 submission-focused closed-loop tests pass. An earlier import-path issue was resolved by running with \texttt{PYTHONPATH=src:.}.

A separate cache implementation recalculates Bayesian posteriors, private evaluations, risk masks, signals, stochastic choices, owner/observer state updates, and first-event endpoints from seeds and saved inputs. All 1,152 trajectory files and 3,840 arms passed, covering 3,815,424 agent-rounds and 487,225 amendment events. Maximum numerical discrepancy was $6.66\times10^{-16}$. The reference deliberately omits declaration identifiers. The 128 synchronized worlds were separately checked for exact equality of every saved individual state and amendment event.

After a compute interruption, the runner checked the completed compressed trajectories and executed only the remaining jobs. The fixed contract, seed blocks, source hashes, and result summaries are preserved. Recorded resume time covers only the resumed segment.

Two internal editorial reviews addressed novelty, identification, and readability. Numerical verification came from the separate reference implementation and new random worlds. The review archive identifies these distinct procedures and their evidence.

\section{V3.1 archived-trajectory analysis}
No new simulations were run for this revision. The analysis script reconstructs every agent's first amendment directly from the event log, verifies the stored endpoints for the strict and all-initial-condemner cohorts, and checks shared private states and review schedules. It processes all 1,152 compressed files (3,840 arms). The record--flow endpoint contrast and its 10,000-draw paired-world bootstrap interval agree with the archived analysis to numerical precision.

\begin{figure}[htbp]\centering
\includegraphics[width=.68\linewidth]{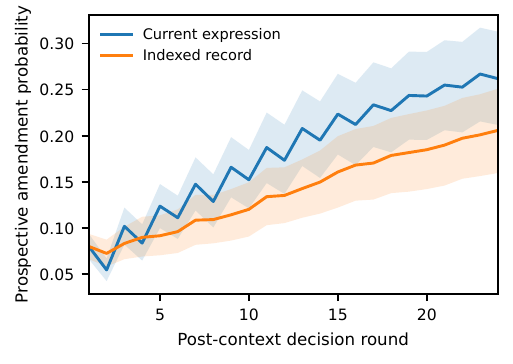}
\caption{Prospective amendment probability for the fixed strict cohort, holding the review-opportunity probability constant. Bands show pointwise bootstrap intervals over worlds.}
\label{fig:propensity}
\end{figure}

The mechanism diagnostic fixes the strict-risk cohort before branching. At each round it averages the observed cue and the prospective amendment probability over that same cohort, including agents who have already amended, then gives each of the 50 eligible worlds equal weight. The plotted prospective probability is $q(1-P_i(+1,t))$: for an author whose original condemnation remains unamended, this is the next-round first-amendment probability. The fixed-cohort curve therefore tracks choice pressure separately from changes in who remains at risk (Figure~\ref{fig:propensity}).

Summing the difference in first-amendment incidence over time exactly recovers the final survival contrast. Decision rounds 1--4, 5--12, and 13--24 account for 4.7435, 12.2043, and 2.4925 percentage points, respectively, summing to 19.4403 before rounding. These are disjoint time bins, not causal mediation shares. The initial matched cue is 0.7812 in both channels. At decision round 8, the mean cues are 0.5593 (flow) and 0.6409 (record), with prospective probabilities 0.1286 and 0.1090. At decision round 24, the respective cue means are $-0.0062$ and 0.2505, with probabilities 0.2619 and 0.2058.

The identifier-free reference implementation reproduced 3,815,424 individual rounds and 487,225 amendments, with maximum absolute discrepancy $6.66\times10^{-16}$. Hash checks also confirmed the unchanged production source and trajectories.

\section{Reproduction commands}
Use Python 3.13 with the provided dependency pins. The bounded revision's checks do not need another simulation run. From the archive's project root:
\begin{Verbatim}[fontsize=\small]
python -m pip install -r requirements-closure.txt
PYTHONPATH=src:. python -m pytest -q
python revision_v31/analysis/reverify_frozen.py
python revision_v31/analysis/analyze_archived.py
python revision_v31/analysis/build_figures.py
cd revision_v31/paper
bash build.sh
cd ../..
python revision_v31/analysis/check_revision.py
\end{Verbatim}
The bundle supplies both BibTeX sources and a pre-generated bibliography. The full computational archive contains the underlying closure trajectories; this LaTeX submission package contains only compilation resources. Results can be regenerated from a separate copy of the frozen archive using the listed commands.

\section{Source access and submission status}
The nearest-neighbour comparison draws on the equation-level audit of private/expressed opinion dynamics, poster-observer interactions, and retraction models. The $\alpha$-EPO citation uses its published Physical Review E entry. Five empirical studies distinguish visible correction responses, self-correction, deletion, diffusion, and mistaken social correction; the source ledger records whether each was consulted in full text or through indexed metadata.

This publicly attributed version retains the V3.1 scientific content, including the Methods disclosure of AI assistance. Experimental code and raw trajectories are retained in the computational archive pending the author's release decision.

\end{document}